\documentclass[aps,pra,floatfix,superscriptaddress,twocolumn]{revtex4-2}

\usepackage{graphicx}

\usepackage{amsmath, amssymb}

\graphicspath{{.}{figures/}{Figures_in_PDF_Format}}

\usepackage[caption=false]{subfig}

\usepackage{braket}

\usepackage[colorlinks=true, allcolors=blue]{hyperref}

\usepackage{tikz}

\begin{document}


\title{Supercharging exceptional points: Full-spectrum pairwise coalescence in non-Hermitian systems}

\author{Yusuf H. Erdoğan}
\affiliation{Department of Physics, Koc University, Istanbul 34450, Turkey}
\author{Masudul Haque}
\affiliation{Institut für Theoretische Physik, Technische Universitat Dresden, 01062 Dresden, Germany}
\begin{abstract}
We consider non-Hermitian tight-binding one-dimensional Hamiltonians and show that imposing a certain symmetry causes all eigenvalues to pair up and the corresponding eigenstates to coalesce in pairs.  This Pairwise Coalescence (PC) is an enhanced version of an exceptional point --- the complete spectrum pairs up, not just one pair of eigenstates.  The symmetry is that of reflection excluding the central two sites, and allowing flipping of non-reciprocal hoppings (``generalized off-center reflection symmetry'').  Two simple examples of PC exist in the literature --- our construction encompasses these examples and extends them to a vast class of Hamiltonians.  We study several families of such Hamiltonians, extend to cases of full-spectrum higher-order coalescences, and show how the PC point corresponds to amplified non-orthogonality of the eigenstates and enhanced loss of norm in time evolution.      

\end{abstract}

\maketitle

\date{\today}

\section{Introduction}

There has been intense interest in non-Hermitian Hamiltonians in recent years.  Although  Hermiticity is a fundamental aspect of standard quantum
mechanics describing closed or isolated quantum systems, non-Hermitian Hamiltonians are useful as effective descriptions of systems where loss or gain plays an important role, such as open quantum systems \cite{Sudarshan_Chiu_Gorini_PRD1978_decaying,  IRotter_JPA2009_opensystems, Ashida_Ueda_AdvPhys2020_review} and optical systems described by wave equations formally analogous to a Schr\"odinger equation \cite{Cao_Wiersig_RMP2015, Ozdemir_Rotter_Nori_NatMaterials2019_review, ElGanainy_etal_CommunicPhys2019_review_nonHermOptics, Ashida_Ueda_AdvPhys2020_review, Parto_Christodoulides_Nanophotonics2021_topologicalphotonics}.

A remarkable aspect of non-Hermitian physics is the appearance of 
exceptional points (EPs), where two or more eigenvalues and their corresponding eigenvectors coalesce   \cite{Kato_book1995, Heiss_Harney_EPJD2001_chirality, Heiss_JPA2004_EPs,  Berry2004, Heiss_JPA2008_chirality, Mueller_IRotter_JPA2008_exceptional, Heiss_JPA2012, Miri_Alu_Science2019_EPs_optics_photonics, Li_EtAl_NatureNanotech2023_EPs_photonics}.   This phenomenon has a large number of nontrivial physical consequences and potential applications, such as control or suppression of lasing \cite{Liertzer_Tuereci_Rotter_PRL2012_lasers, Fischer_EtAl_NatureNanophotonics2024_controlling_lasing}, vanishing group velocity of light \cite{Goldzak_Moiseyev_PRL2018_LightStops}, enhanced sensitivity to perturbations and the use of this phenomenon for sensing \cite{Wiersig_PRL2014_sensitivity_EP, Wiersig_PRA2016_sensors_at_EP, Hodaei_ElGanainy_Christodoulides_Khajavikhan_Nature2017_enhancedsensitivity, Chen_Ozdemir_Wiersig_Yang_Nature2017_sensing_microcavity, Mao_Yang_ScienceAdvances2024_phasesensing}, unidirectional transmission or propagation \cite{Lin_Kottos_Christodoulides_PRL2011_unidirectional, Regensburger_etal_NatPhys2012, Yin_Zhang_NatureMaterials2013_commentary_unidirectional, WuArtoniRocca_PRL2014_unidirecional, Shi_etal_NatComm2016_acoustics_unidirectional} and other chiral phenomena \cite{Heiss_Harney_EPJD2001_chirality, Dembowski_Harney_Heiss_Richter_PRL2003_expt_microwavecavity, Heiss_JPA2008_chirality, Zhu_EtAl_OpticsLett2013_invisiblecloak, Fleury_Sounas_Alu_NatComm2015_acousticsensor, Oezdemir_Wiersig_Rotter_PNAS2016_chiralmodes, Sui_EtAl_OptLett2016_unidirectional_emission, Miao_Longhi_Litchinitser_Feng_Science2016_microlaser, Shen_Cummer_PRMater2018_accoustic,  Gao_Snoke_Ostrovskaya_PRL2018_excitonpolariton_chiral, Zhang_EtAl_PRX2018_encircling_EP, Baek_EtAl_Light2023_graphenemetasurfaces_chiral}, and a rapidly growing number of topological effects \cite{Lee_PRL2016_AnomalousEdgeState, Kozii_LiangFu_arxiv2017_quasiparticles, Leykam_Nori_PRL2017_edgemodes, Lieu_PRB2018_nonHermSSH, Zhong_Christodoulides_ElGanainy_NatComm2018_winding, Yao_Wang_PRL2018_EdgeStates_TopologicalInvariants, Shen_Zhen_LiangFu_PRL2018_TopologicalBandTheory, Kunst_Edvardsson_Budich_Bergholtz_PRL2018_biorthogonal, Budich_Kunst_Bergholtz_PRB2019_nodalphases, BergholtzBudichKunst_RMP2021, Li_EtAl_NatureNanotech2023_EPs_photonics, Okuma_Sato_AnnuRevCondMat2023_nonhermtopology, Yang_Koenig_Bergholtz_RepProgPhys2024_homotopy}. 
Higher-order EPs, where more than two eigenstates coalesce, have enjoyed significant interest \cite{Graefe_Korsch_Niederle_JPA2008_BH_higherorderEPs, Demange_Graefe_JPA2011_threefoldEP, Hodaei_ElGanainy_Christodoulides_Khajavikhan_Nature2017_enhancedsensitivity, Nada_Othman_Capolino_PRB2017_higherorder,  Jing_Oezdemir_Nori_SciRep2017_higherorder_optomechanics, Pan_Cui_PRA2019_higherorder_Bosegases, Zhang_You_PRB2019_cavitymagnonics, Zhang_Zhang_Song_PRA2020_higherorder_supersymmetric, Yu_Yang_Cao_PRB2020_higherorder_ferromagnetictrilayers, Mandal_Bergholtz_PRL2021_higherorder, Wang_Guo_Berakdar_PhysRevAppl2021_higherorder_magnonicwaveguides, Wiersig_PRA2022_higherorder_hierarchical, Grom_arxiv2024_higherorder, Montag_Kunst_PRR2024}, partially because of the promise of enhancing the interesting and possibly useful features of an EP.

In this work, we devise and study a potentially more potent enhancement of the EP phenomenon.   We present a mechanism for generating \emph{full-spectrum} pairwise coalescence (PC), such that we have not just two eigenstates pairing up and coalescing, but rather all eigenstates (the \emph{complete} spectrum) coalescing pairwise, at the same parameter value.  Our setting is a tight-binding Hamiltonian on a one-dimensional chain with an even number ($L=2k$) of sites.   We present an extensive class of Hamiltonians and prove that every member of this class displays PC: at a particular parameter value, there are only $k=L/2$ distinct eigenvalues and $k=L/2$ linearly independent eigenstates.  The full set of Hamiltonians incorporates many different types of physics: we examine in particular a few families of Hamiltonians supporting PC.       

Our scheme involves imposing reflection symmetry on the chain excluding the central two sites (``off-center reflection symmetry''), and then tuning the $2\times2$ Hamiltonian of the central two sites to an EP.  Additionally, a set of exchanges in the hoppings are allowed; so we describe the symmetry as ``generalized off-center reflection symmetry''.   We also expand our strategy to design classes of Hamiltonians with 4-fold (or 8-fold or in general $2^n$-fold) coalescences throughout the spectrum.  

Regular exceptional points --- a coalescence of just two eigenstates --- have physical consequences and signatures.  We expect that pairwise coalescence in the complete spectrum will have various signatures that are more pronounced compared to regular EPs.  We briefly explore two effects.  The first is a dramatic enhancement of the non-orthogonality of eigenstates.  The lack of orthogonality of eigenstates is a hallmark of non-Hermitian Hamiltonians.  (The eigenstates  possess bi-orthogonality instead of regular orthogonality.)  This lack of orthogonality is pronounced at an EP for which two eigenstates coalesce, i.e., are maximally non-orthogonal.  In our case, we show that this effect is manyfold enhanced as the full spectrum (not just one pair) is paired up.  The second signature of PC examined in this work concerns the loss of wavefunction norm in time evolution.  Since time evolution is not unitary under a non-Hermitian Hamiltonian, the norm is not conserved.  Focusing on systems where there is only absorption (wo that the norm only decreases and does not increase),  we show that the loss of norm at long times is maximal near the PC point.

\begin{figure}[tbp]
\centering
\includegraphics[width=8.6cm]{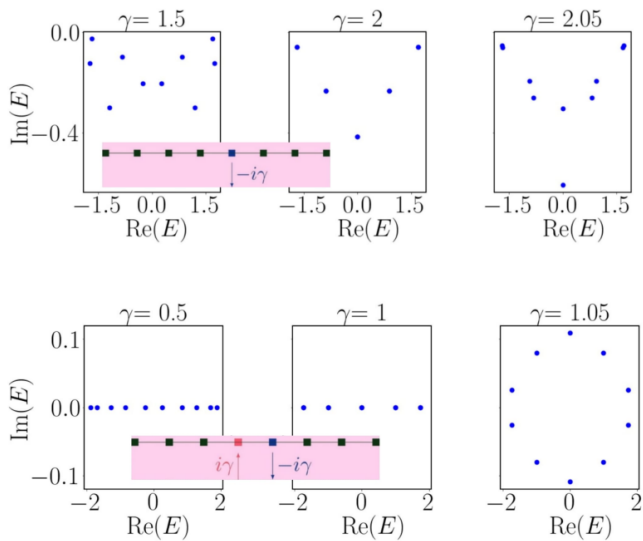}
\caption{Demonstration of PCs known in the literature. Top:  $\alpha=0$ in Eq.\ \eqref{eq_previous_cases}; Ref.\ \cite{Burke2020}.  All eigenvalues pair up (PC occurs) at $\gamma=2$.   
Bottom: $\alpha=-\gamma$; Ref.\ \cite{Ortega2020}. PC is seen at $\gamma=1$.  Spectrum is shown for $L=10$ sites in both cases so that there are 10 eigenvalues but in the central panels only 5 are visible. 
Insets show the model geometries schematically, with arrows indicating loss or gain. 
}
\label{fig_previous_cases}
\end{figure}

Our study of PCs was motivated partly by some results in Ref.\ \cite{Burke2020}, where PC was numerically found and analytically proven in an even-sized chain with a negative imaginary potential on one of the central sites.  PC can also be seen in the spectral data presented in Ref.\ \cite{Ortega2020}, which considers a PT-symmetric tight-binding chain, with imaginary potentials of opposite signs in the two central sites. 
Defining $L=2k$, we can express both previously studied Hamiltonians as 
\begin{equation}
\label{eq_previous_cases}
    H = -\sum_{i=1}^{L-1} \ket{i} \bra{i+1} + h.c. - i \alpha \ket{k} \bra{k} - i \gamma \ket{k+1} \bra{k+1}. 
\end{equation}
The model in Ref.\ \cite{Burke2020} corresponds to  $\alpha=0$, and shows PC at $\gamma=2$.  The $\mathcal{PT}$-symmetric Hamiltonian of Ref.\ \cite{Ortega2020} corresponds to $\alpha=-\gamma$, and shows PC at $\gamma=1$.  

In Figure \ref{fig_previous_cases}, we present the spectrum of both these non-Hermitian Hamiltonians at values of $\gamma$ below, at, and above the value at which the EP's appear ($2$ or $1$).  The phenomenon of PC is illustrated by the observation that only half the eigenvalues are visible at that value of $\gamma$.

We will follow the practice of Eq.\ \eqref{eq_previous_cases}: we set $\hbar=1$ and set (one of) the hoppings to be unity whenever possible.  This sets the units of time and energy in all our systems. Accordingly, explicitly specifying units is not required for any of this manuscript's axis/figure captions.

In Section \ref{sec:mainresult} we explain in detail the generalized off-center reflection symmetry.  We then present the class of Hamiltonians which we have found to display PC via this symmetry.  
In Section \ref{sec:example_families} we list several example families out of this vast class of Hamiltonians.  
In Section \ref{sec:mainproof} we prove our claim mathematically.  Readers who want to focus on the physical results and examples should be able to follow the rest of the paper without pursuing the details of this section. 
Sections \ref{sec:nonorthogonal} and \ref{sec:normloss} discuss two consequences of PC: in Section \ref{sec:nonorthogonal} we analyze the pronounced non-orthogonality at the PC point and in Section \ref{sec:normloss} we present the substantially enhanced loss of norm (in time evolution) at and near PC.  
Finally, we provide some outlook and context in Section \ref{sec:discussion}.
Appendix \ref{appsec:derivation_ATrelation} contains the derivation of an intermediate result used in Section \ref{sec:mainproof}.

\section{Off-center reflection symmetry and Pairwise Coalescence \label{sec:mainresult}}

We now present the general class of tight-binding Hamiltonians that we have found to possess pairwise coalescence.  The basic idea is to impose reflection symmetry excluding the central two sites (off-center reflection symmetry), and then tune the central two sites to an exceptional point.   In this section, we explain this general scheme and introduce the structure of the resulting Hamiltonian.  In the following section (Section \ref{sec:example_families}) we will consider specific example families following from this general structure.  The analytic proof of PC is given in Section \ref{sec:mainproof}.


We consider tight-binding nearest-neighbor chains of length $L=2k$, where $k$ is an integer.   We will use tight-binding one-dimensional Hamiltonians that have reflection symmetry except for the central $2\times2$ matrix and require the central $2\times2$ matrix to have an EP.  

\begin{figure}[tbp]
\centering 
\includegraphics[width=\columnwidth]{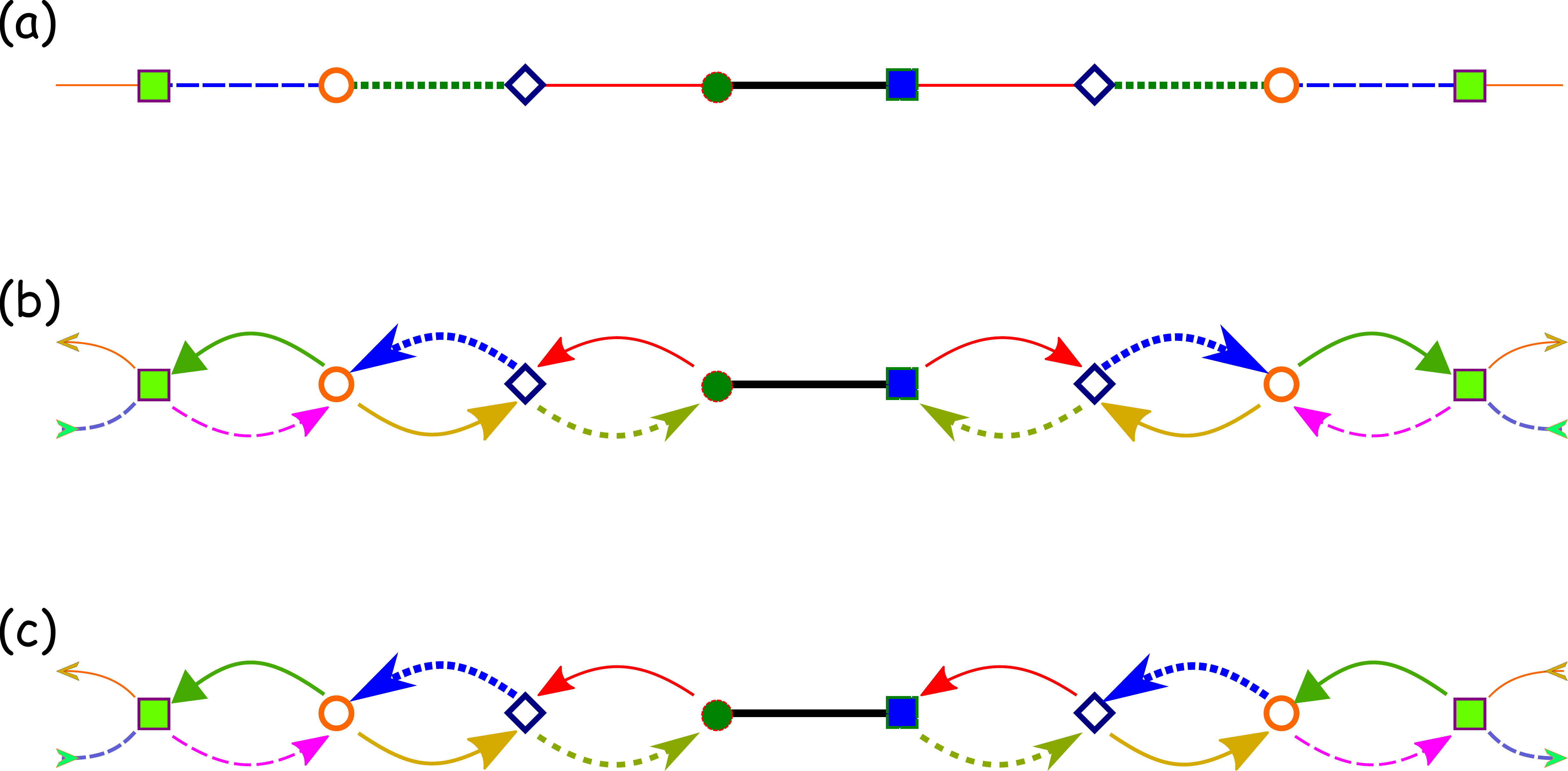}
\caption{Schematic of a tight-binding non-Hermitian chain with (generalized) off-center reflection symmetry. (a) For pedagogy, we first demonstrate the special case with symmetric hoppings on each bond, represented by a single line for each bond.  All bond hoppings are reflection-symmetric.  The on-site terms are all reflection-symmetric except for the central two sites.   
(b) Since the chain is non-Hermitian, the leftward and rightward hoppings on each bond are generally different.  Off-center reflection symmetry requires the leftward (rightward) hopping on the $(j,j+1)$ bond to be the same as the rightward (leftward) hopping on the $(L-j+1,L-j)$ bond.  
(c) Our construction also holds if we flip the leftward and rightward hopping matrix elements on an arbitrary number of bonds.  Here we have flipped all the bonds on the right half of the chain.  This is the version used in Eq.\ \eqref{eq:Ham_general}. 
}
\label{fig:schematic_offcentersym}
\end{figure}

The Hamiltonian matrix is tridiagonal, and off-center reflection symmetry implies   
\begin{equation}
H_{j,j+1}=H_{L-j+1,L-j}  \quad \text{for} \quad  j = 1,\ldots, k-1. 
\end{equation}
This symmetry is illustrated schematically in Figure \ref{fig:schematic_offcentersym}(a,b).  
For ease of description, we first show [Schematic (a)] a case with symmetric real hopping on every bond.  Schematic (b) shows the case of imbalanced bond hoppings while preserving the off-center reflection symmetry.  These hoppings can be real or imaginary or complex. 

Our scheme is somewhat more general: starting from the off-center reflection symmetric case we can exchange the left and right hopping terms on any or all of the bonds; the pairwise coalescence survives such exchanges.  One such case is shown in Schematic (c) of Figure \ref{fig:schematic_offcentersym}: all the bonds on the right half are flipped.  We will use this case, 
\begin{equation}
H_{j,j+1}=H_{L-j,L-j+1}  \quad \text{for} \quad  j = 1,\ldots, k-1, 
\end{equation}for definiteness.  In Section \ref{sec:mainproof} it will be clear that our proof of PC allows one to choose any number of such exchanges.

The onsite potentials can be real or imaginary complex; the off-center potentials (i.e., those on all but the two central sites) are constrained to be reflection symmetric, as indicated by the shapes with which the sites are represented in  Figure \ref{fig:schematic_offcentersym}.

We claim that when the central $2\times2$ matrix has an exceptional point, a lattice with (generalized) off-center reflection symmetry will display pairwise coalescence.  For concreteness, we focus on the case that the central bond has symmetric real hopping and the central two sites have imaginary on-site potentials.  (This restriction is convenient but is not necessary at all, see \S\ref{subsec:generalizing_central_block}.)  With this choice, the
central $2\times2$ matrix has the form  
\begin{equation} \label{eq_centralblock_Ham}
M = \begin{bmatrix}
    H_{k,k} & H_{k,k+1} \\
    H_{k+1,k} & H_{k+1,k+1}
\end{bmatrix}
= - \begin{bmatrix}
    i\alpha & \delta \\
    \delta & i\gamma
\end{bmatrix} ,   
\end{equation}
with real $\alpha$,  $\gamma$,  $\delta$.  
To get equal eigenvalues, we impose that the characteristic polynomial $\det(M-\lambda I)$ has the form $(\lambda-Z)^2$, where $\lambda$ is the eigenvalue variable and $Z$ is an arbitrary complex number.  This leads to  
\begin{equation}  \label{eq:central_block_condition}
    \delta = \pm \frac{\gamma-\alpha}{2}, \quad \text{or} \quad \gamma = \alpha \pm 2\delta.
\end{equation}
Combining the off-center and central parts described above, we obtain the class of tight-binding chain Hamiltonians having the form 
\begin{multline}  \label{eq:Ham_general}
H =  -\sum_{i=1}^{k-1} \Big( b_i \ket{i} \bra{i+1} + c_i \ket{i+1} \bra{i} \Big)
\\   
     - \sum_{i=k+2}^{L} \Big( b_{L-i+1} \ket{i} \bra{i+1} + c_{L-i+1} \ket{i+1} \bra{i}\Big) 
     \\ 
 -\sum_{i=1}^{k-1} a_i  \ket{i} \bra{i}   -\sum_{i=k+2}^{L} a_{L-i+1}  \ket{i} \bra{i}  
     \\
   - \delta \left( \ket{k} \bra{k+1} + h.c. \right) 
   \\ 
   - i \alpha \ket{k} \bra{k} - i \gamma \ket{k+1} \bra{k+1} , 
\end{multline}
with arbitrary complex $a_i$, $b_i$, $c_i$ and real $\alpha$, $\gamma$ and $\delta$ that are related by Eq.\ \eqref{eq:central_block_condition} but otherwise arbitrary.  The Hamiltonian has generalized off-center reflection symmetry, in the sense described above.  (If we swapped the coefficients $b_{L-i+1}$ and $c_{L-i+1}$ in the second line, we would obtain exact off-center reflection symmetry.)  

PC is obtained when the condition \eqref{eq:central_block_condition} is used to fix $\delta$ as a function of $\gamma$ and $\alpha$.     This condition still leaves us with a very large class of Hamiltonians.   
This entire class of Hamiltonians has pairwise coalescence in the full spectrum.  This is the central result of this work.  Our proof of this statement is somewhat lengthy and hence deferred to Section \ref{sec:mainproof}.  

In our numerical demonstrations, we will usually fix  $\alpha$ and $\delta$ and vary $\gamma$, to show that PC occurs when $\gamma=\alpha\pm2\delta$. 

\begin{figure}[tbp]
\centering \includegraphics[width=8.6cm]{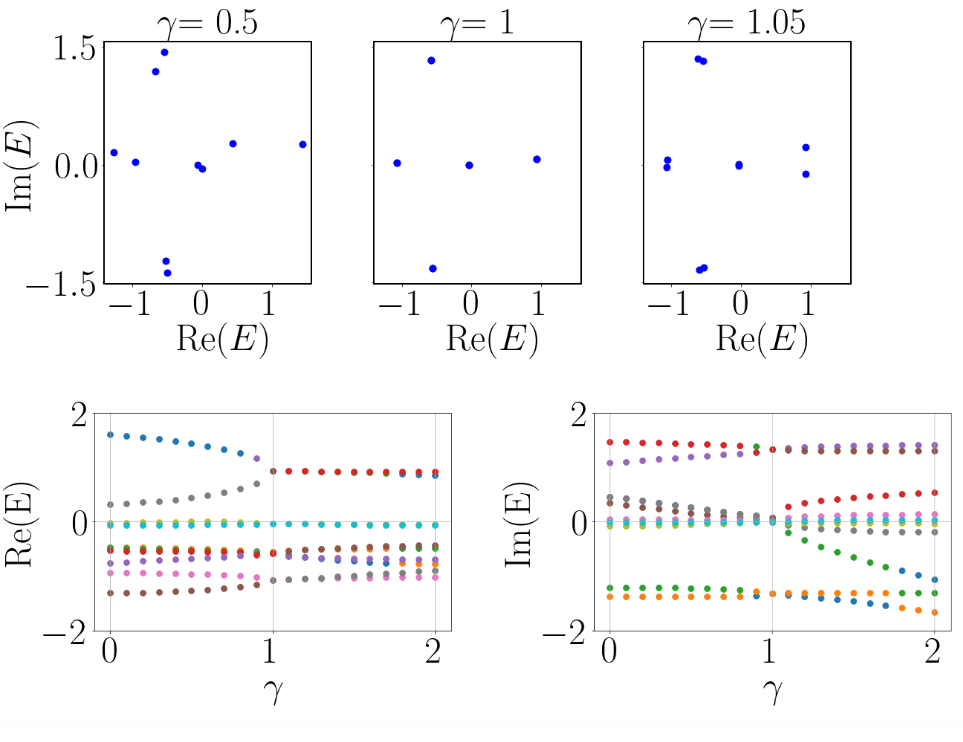}
\caption{Demonstration of PC for a Hamiltonian of form (\ref{eq:Ham_general}) with random coefficients.  The $a_i$, $b_i$, $c_i$ coefficients are of the form $\sigma+i\eta$ where $\sigma$, $\eta$ are sampled from $\mathcal{N}(0,1)$.  We use $L=10$ sites, $\alpha=-1.2,  \delta=-1.1$, so that there should be a PC at $\gamma=1$.  Indeed, there are only 5 visible eigenvalues at $\gamma=1$ (top row), and the real and imaginary parts of the eigenvalues can be seen to pair up at $\gamma=1$ (bottom row).}

    \label{fig_random_matrixelements}
\end{figure}

As a very generic demonstration, we choose the matrix elements $a_n,b_n$ and $c_n$ to be complex and treat both real and imaginary parts as Gaussian random variables,  drawing both from the normal distribution $\mathcal{N}(0,1)$.  Figure \ref{fig_random_matrixelements} demonstrates PC for a particular realization.  In the displayed case, we have used $\alpha=-1.2$ and $\delta=1.1$, and we vary $\gamma$.  Eq.\ \eqref{eq:central_block_condition}  predicts that PC should occur at $\gamma=\alpha+2\delta=1$ for these parameters.  Indeed, in Figure \ref{fig_random_matrixelements} we see this both in the plots of eigenvalues against $\gamma$ (top panels), and in the complex-plane scatter plots of the eigenvalues (bottom panels).

Why does the combination of off-center reflection symmetry and central EP produce pairwise coalescence in the entire spectrum?  The algebraic proof of PC (Section \ref{sec:mainproof}) does not immediately provide intuition.  We found it helpful to think of the limit where the central block is ``switched off'', i.e., $M$ of Eq.\ \eqref{eq_centralblock_Ham} is replaced by a zero matrix.  In this case, the system consists of two chains that are identical upon reflection, so that the spectrum is composed of $L/2$ degenerate pairs.  Switching on the central block $M$ lifts this degeneracy in general.  It turns out that, if the spectrum of $M$ is itself defective, i.e., if $M$ has an EP, then the degeneracy is not lifted, so that the entire spectrum is arranged as $L/2$ coalesced pairs.

\section{Notable families showing PC \label{sec:example_families}}

\begin{figure}[tbp]
\centering \includegraphics[width=\columnwidth]{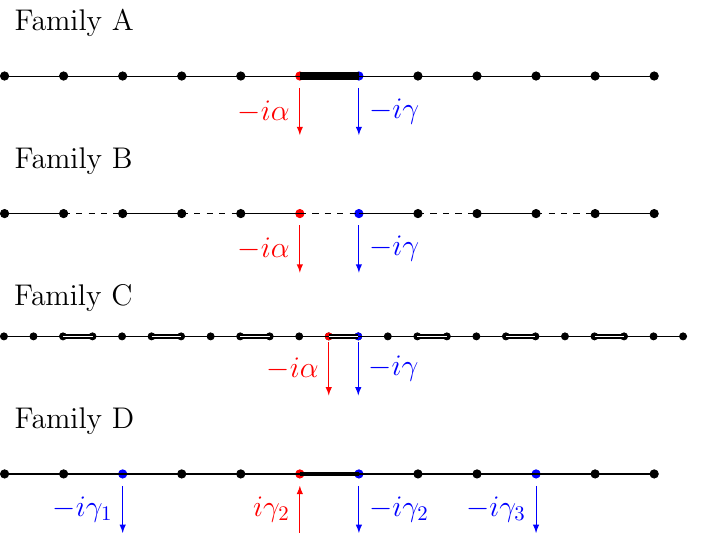}
\caption{ \label{fig:schematics_severalmodels}
Schematic of several subclasses of the general class described in Eq.\ \eqref{eq:Ham_general}.  
Family A:  The central bond has arbitrary hopping, and the central two sites have imaginary potentials, in an otherwise uniform lattice.  PC can be obtained by tuning $\alpha$, $\gamma$ and the central hopping. 
Family B:  A nonhermitian version of the SSH Hamiltonian.  As in Family A, the central bond hopping and central site potentials are tuned.  The other hopping strengths alternate in strength.
Family C: We choose periodically alternating bonds with period three and place imaginary potentials on the central sites, Eq.\ \eqref{eq:periodthreelattice}.
Family D: $\gamma_1$ is at $L/4$ and $\gamma_3$ is at $3L/4+1$. This Family has PC for $\gamma_1=\gamma_3$ and 4-wise coalescence for $\gamma_1=2\gamma_2=\gamma_3=2$. 
}
\end{figure}

In this section, we consider several families of Hamiltonians displaying PC, which are all part of the broad class of Hamiltonians \eqref{eq:Ham_general} introduced in the previous section.   First, in \S\ref{subsec_previous_cases} we point out that the simple models for which PC appeared previously in the literature \cite{Burke2020, Ortega2020} are special cases of Eq.\ \eqref{eq:Ham_general}.  We wish to go beyond these simplest models, but we also would like to restrict to models that are more physically plausible than the extreme `random' case of Figure \ref{fig_random_matrixelements}.  In Figure \ref{fig:schematics_severalmodels} we show schematically several other families of tight-binding chains that display the PC phenomenon.  We describe them in the following subsections. \S\ref{subsec_FamilyA} discusses minimal generalizations of the previously known cases (Family A of Figure \ref{fig:schematics_severalmodels}). \S\ref{non_hermitian_ssh_subsec} introduces non-Hermitian variants of the celebrated SSH chain displaying PC (Family B and Family C).  In \S\ref{subsec_higher_order}  we show how our scheme can be extended to produce Hamiltonians with higher-order EP clusters, i.e.,  non-Hermitian models for which the eigenvalues all group together not just in pairs, but into quadruplets, or larger groups (Family D).

\subsection{Previously known cases \label{subsec_previous_cases}}

The two cases previously known \cite{Burke2020, Ortega2020} are relatively simple examples of the class of Hamiltonians \eqref{eq:Ham_general}.  They both correspond to symmetric and uniform hoppings, $b_i=c_i=\delta=1$, with no on-site potentials outside the central block.   These conditions reduce Eq.\ \eqref{eq:Ham_general} to Eq.\ \eqref{eq_previous_cases}.  
 
Ref.\ \cite{Burke2020} deals with a single loss potential, $\alpha=0$, $\gamma>0$.  According to our condition for the central block, Eq.\ \eqref{eq:central_block_condition}, this model has PC at $\gamma=2$.  This is as reported and analyzed in Ref.\ \cite{Burke2020}, and shown in Figure \ref{fig_previous_cases} (top row) of this Article.  

The case of Ref.\  \cite{Ortega2020}  is $\mathcal{PT}$-symmetric, with a loss and a gain in the central block: $\alpha =-\gamma$.  Our condition \eqref{eq:central_block_condition} shows that this will have PC at $\gamma=1$, as shown in Figure \ref{fig_previous_cases} (bottom row).  Although PC is not explicitly discussed in   Ref.\  \cite{Ortega2020} as a general phenomenon, it can be seen at $\gamma=1$ by examining Figure 2 of that work.  
The value $\gamma=1$ also happens to be the  $\mathcal{PT}$-symmetry breaking point.

\subsection{Minimal extensions of previous cases \label{subsec_FamilyA}}

We now consider some straightforward extensions of the previously observed cases  \cite{Burke2020, Ortega2020}, which are represented above by Eq.\ \eqref{eq_previous_cases}.  A natural generalization is to modify only the central block, keeping the off-center hoppings uniform ($=1$) and the off-center sites potential-free.  With imaginary potentials $-i\alpha$ and $-i\gamma$ on the two central sites, we will obtain a PC when the central hopping is $\frac{1}{2}(\gamma-\alpha)$.  This is the geometry shown as Family A in Figure \ref{fig:schematics_severalmodels}. 

\begin{figure}[tbp]
    \centering
\includegraphics[width=\columnwidth]{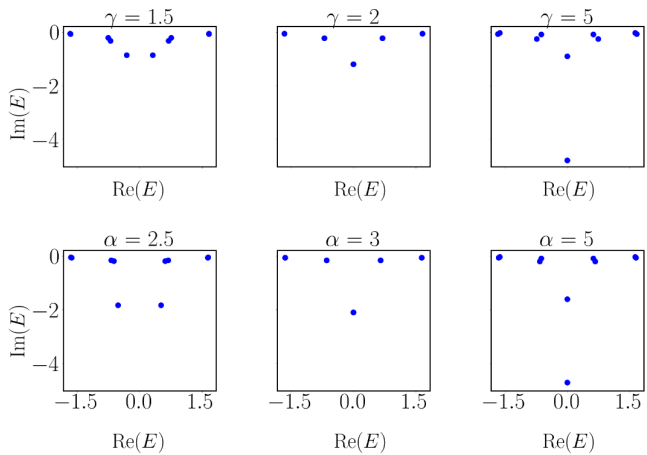}
\caption{   \label{fig_modelA_demo}
Illustration of PC in Family A, Eq.\ \eqref{Ham_Family_A}.  Top: we let $\alpha=1.5, \delta=-0.25$ so that PC happens at $\gamma=2.$  Bottom:  we use $\gamma=2,\delta=0.5$ so that PC happens at $\alpha=3$.  
}   
\end{figure}

With $L=2k$, the Hamiltonian of Family A is 
\begin{equation}
\label{Ham_Family_A}
H_{A} =  H_{A,\text{offc}} + H_{A,\text{cent}} 
\end{equation}
with the off-center part 
\begin{multline}
\label{Ham_Family_A_o}
    H_{A,\text{offc}} =  -\sum_{j=1}^{k-1} \Big( \ket{j} \bra{j+1} + \ket{j+1}\bra{j}\Big)
      \\  - \sum_{j=k+1}^{L-1} \Big( \ket{j} \bra{j+1} + \ket{j+1}\bra{j} \Big)
\end{multline}
and central part 
\begin{multline}
\label{Ham_Family_A_c}
H_{A,\text{cent}} =  - \delta \Big( \ket{k} \bra{k+1} + \ket{k+1}\bra{k} \Big) 
\\
- i \alpha \ket{k} \bra{k} - i \gamma \ket{k+1} \bra{k+1} .
\end{multline}
The previous cases  \cite{Burke2020, Ortega2020} display PC only at one particular finely tuned value of parameters.  In contrast, our Family A is a three-parameter $(\alpha,\gamma,\delta)$ family of Hamiltonians, which displays PC on two two-dimensional sub-manifolds of parameters, obtained by imposing the conditions  $\delta=(\gamma-\alpha)/2$ and  $\delta=-(\gamma-\alpha)/2$.  We demonstrate PC in this family in Figure \ref{fig_modelA_demo}, with the central bond chosen to have different hopping compared to the bulk of the chain.


In addition, one can symmetrically add pairs of potentials to the two edges of the chain.  
Motivated by a long-standing interest in Hermitian boundary impurity models \cite{Boundary1,Boundary2,Boundary3}, and more recent interest in non-Hermitian boundary impurity models \cite{Zhu_Chen_PRA2014_nonHermSSH, Wiersig_PRA2018_nonorthogonality,NonHermitianBoundary2_2021,NonHermitianBoundary3_2021, NonHermitianBoundary1_2023}, we point out that we can maintain pairwise coalescence if we set  $H_{1,1}=H_{L, L} = -i\beta$:
\begin{equation}
H = H_{A} - i\beta\ket{1}\bra{1} - i\beta\ket{L}\bra{L} 
\end{equation}
for any $\beta$, i.e., if we add arbitrary equal potentials at the two edges of the chain.  Here $\beta$ can be real (imaginary potentials), imaginary (regular potentials), or complex (a combination of the two).


\subsection{Non-Hermitian variants of SSH Hamiltonian \label{non_hermitian_ssh_subsec}
}

The SSH (Su-Schrieffer-Heeger) Hamiltonian \cite{Su_Schrieffer_Heeger_PRL1979} is a foundational model of topological condensed matter theory.  In recent years, a number of non-Hermitian variants of the SSH Hamiltonian have appeared in the literature, involving either non-reciprocal hoppings \cite{Lieu_PRB2018_nonHermSSH, Yin_Jiang_Chen_PRA2018_nonHermSSH_windingnumber, Yokomizo_Murakami_PRL2019, Chen_PhysRevB_2019_nonHermSSH,Herviou_PhysRevA_2019_nonHermSSH,Flebus_Duine_Hurst_PRB2020_nonHermSpinTorque,Arouca_PhysRevB_2020_nonHermSSH,He2021_nonHermSSH, Fleckenstein_PhysRevRes_2022_nonHermSSH,Pyrialakos_PhysRevLetters_2022_nonHermSSh,Gunnik_Duine_PRB2022_nonHermSSH,Aquino_PhysRevB_2023_nonHermSSH, Halder_Basu_JPCM2023_nonHermSSH, Li_Zhang_Gong_PRB2024_nonHermSSH,Lin_nonHermSSH_PhysRevB_2024}  or imaginary on-site potentials \cite{Zhu_Chen_PRA2014_nonHermSSH,Yuce_nonHermSSH2018,Yuce_Oztas_SciRep2018_nonHermTopol, Dengel_Wunner_PRA2018_nonHermSSH, Flebus_Duine_Hurst_PRB2020_nonHermSpinTorque,WU_nonHermSSH2021,Garmon_Noba_PRA2021_nonHermSSH_PTsym, Ostahie_Aldea_PhysLett2021_nonHermSSH, Gunnik_Duine_PRB2022_nonHermSSH, Ezawa_nonHermSSH2022,Chang_PhysRevA_2023_nonHermSSH,Slootman_Bergholtz_Morais_PRR2024_nonHermSSH,SHI_nonHermSSH_2024,Rottoli_2024_nonHermSSH} or both. Here, we introduce variants of the SSH chain which support pairwise coalescence.  We will refer to this class of models as Family B.  

 We define the SSH Hamiltonian (without non-Hermitian modifications) as
\begin{equation} \label{eq:ssh_bare}
H_{\text{SSH}} = - \sum_{j=1}^{L/2} \left( J_1 \ket{2j-1}\bra{2j} + J_2 \ket{2j}\bra{2j+1} + \text{h.c.} \right).
\end{equation}
As usual, we consider only even $L=2k$.  We now place imaginary potentials $-i\alpha$ and $-i\gamma$ at the central sites:
\begin{equation} \label{eq:familyB}
 H_B =   H_{\text{SSH}}  - i \alpha \ket{k} \bra{k} - i \gamma \ket{k+1} \bra{k+1} .
\end{equation}
For even $k$ ($L$ is divisible by $4$, i.e., $L=4m$ with integer $m$), the central bond is a $J_2$ bond.  Figure \ref{fig:schematics_severalmodels} shows an example in this class.  In this case, we obtain PC for $\gamma-\alpha = 2J_2$.  The PC holds for arbitrary $J_1$.  For odd $k$ ($L=4m+2$ with integer $m$), the central bond is a $J_1$ bond.   PC now appears for $\gamma-\alpha = 2J_2$, for arbitrary $J_2$.  

\begin{figure}[tbp]
\centering
\includegraphics[width=8.6cm]{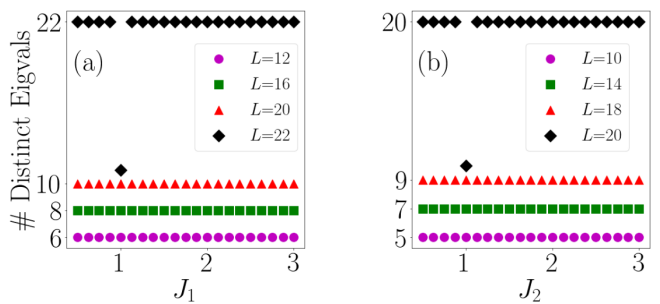}
\caption{Demonstration of PCs in Family B, with $\alpha=0$. (a) We fix $J_2=1$ and $\gamma=2J_2=2$.  Then PC is observed when $L$ is a multiple of $4$ ($L=4m$, exemplified here by $L=12$, $16$ and $20$).  In these cases, the number of distinct eigenvalues is seen to be $L/2$ for any $J_1$.  In the $L=22$ case, PC is not seen (there are 22 distinct eigenvalues) except for the special case $J_1=J_2=1$.  (b) We use $J_1=1$ and $\gamma=2J_1=2$.  PC can be seen for $L=4m+2$, at arbitrary $J_2$: the number of distinct eigenvalues is $L/2$ instead of $L$.  For $L=4m$ (exemplified by $L=20$), there is no PC except at $J_1=J_2$.
We take two eigenvalues to be distinct if their magnitude differs by more than $10^{-5}$.}
\label{fig_SSH_numDistinct} 
\end{figure}

The phenomenology is demonstrated in   Figure \ref{fig_SSH_numDistinct}, where we have used only one imaginary potential, $\alpha=0$.  In panel (a), we use $\gamma=2J_2$.  Then for $L=4m$, we should see PC for any $J_1$; indeed we see only $k=L/2$ distinct eigenvalues for $L=12$, $16$, and $20$.  For $L=4m+2$, we have no PC in general, and there are $L$ distinct eigenvalues, as shown for $L=22$.  The exception is the case $J_1=J_2$, in which case we recover the model of Ref.\ \cite{Burke2020}, i.e., Eq. \eqref{eq_previous_cases} with $\alpha=0$ and $\gamma=2J_1=2J_2$.  An analogous situation is seen for $\gamma=2J_1$ (Panel (b) of Figure \ref{fig_SSH_numDistinct}), with the behaviors of $L=4m$ and $L=4m+2$ now reversed. 

In principle, we can also modify the central bond strength (vary $\delta$) and construct a larger class of models with PC, for the price of losing the exact periodicity of the hopping strengths.  We will not pursue this direction here.

Variants with more involved periodic hopping patterns are also possible.  One example is based on the period-3 chain:
\begin{multline}
\label{eq:periodthreelattice}
H = - \sum_{j=1}^{L/3-1} \Big( J_1 \ket{3j-2}\bra{3j-1} + J_1 \ket{3j-1}\bra{3j}  \\
    + J_2 \ket{3j}\bra{3j+1} \Big) + \text{h.c.}
\end{multline}
As in the SSH case, we can now add impurities to the central two sites.  Off-center reflection symmetry is present when $L$ is a multiple of $6$, with the central bond strength then being $\delta=J_3$.  PC is obtained at $\gamma=\alpha\pm 2J_3$.

\subsection{Higher Order Exceptional Points \label{subsec_higher_order}}

This work concerns pairwise coalescence of the entire spectrum: all $L$ eigenstates pairing up so that we have $L/2$ pair coalescences, not just one pair coalescence as in a usual EP.  It would be even more remarkable if we could make the entire spectrum separate into groups larger than 2 which each coalesce.  Here, we show how we can extend our scheme to have the $L$ eigenvalues organized into $L/4$ quadruplets, with the eigenstates of each group undergoing a 4-wise coalescence.  
There has been significant interest in higher-order EPs (as noted in the Introduction), but we are not aware of other examples where the \emph{full spectrum} breaks up into higher-order groups of coalescent eigenvalues.

Our Hamiltonians with PC are constructed by connecting together two halves whose $\frac{L}{2}\times\frac{L}{2}$ Hamiltonians have identical spectra due to off-center reflection symmetry.  We can now arrange the two identical $\frac{L}{2}\times\frac{L}{2}$ Hamiltonians themselves to be degenerate, i.e., to have PC.  Then, when they are joined by a central part containing an EP, the resulting $L\times{L}$ spectrum consists of 4-fold coalesced clusters.  

\begin{figure}[htbp]
\centering 
\includegraphics[width=\columnwidth]{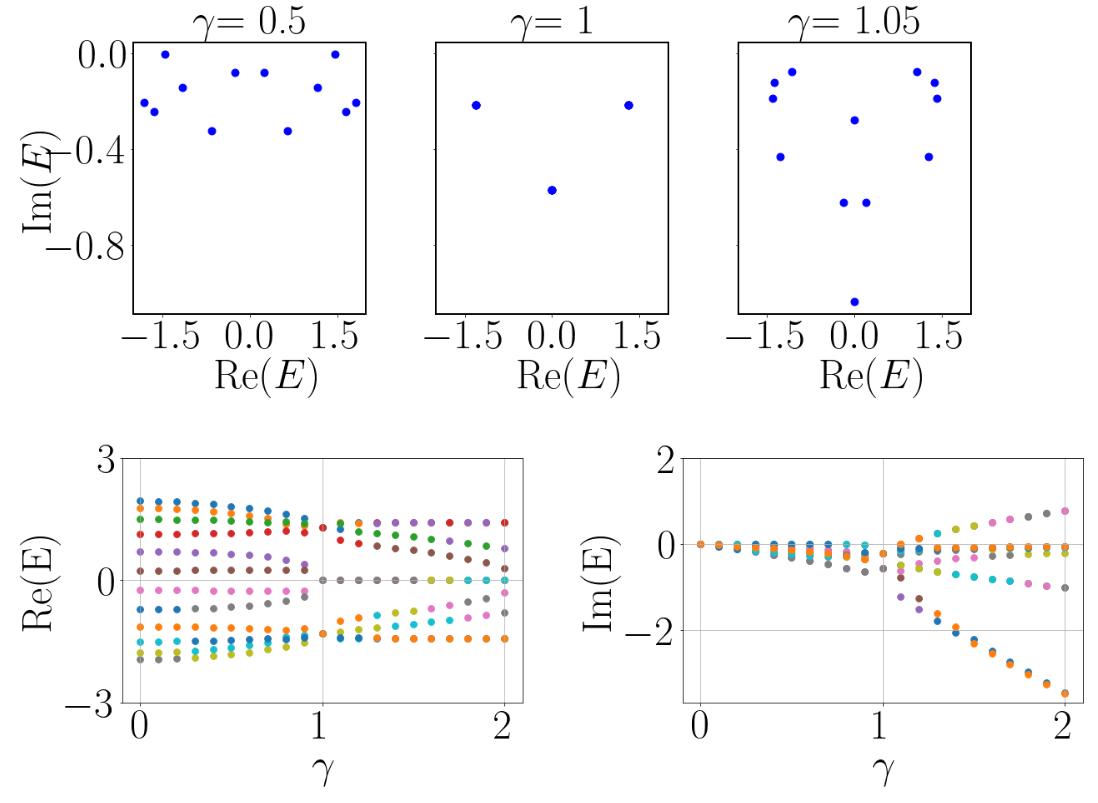}
\caption{Demonstration of 4-wise coalescence in Family D, Eq.\ \eqref{eq:familyD}, for an $L=12$ chain with $\gamma_1/2=\gamma_3/2=\gamma_2=\gamma$.  4-wise coalescence implies that there exists $L/4$ different eigenvalues for an $L\times L$ matrix.  At the 4-wise coalescence point $\gamma=1$, there are $m=3$ distinct eigenvalues (3 different real parts and 2 different imaginary parts).}
    \label{fig_higher_order}
\end{figure}

As an example, we have Family D in Figure \ref{fig:schematics_severalmodels} with $L=4m$:
\begin{equation} \label{eq:familyD}
\begin{aligned}
    H_D =&  -\sum_{\substack{i=1 \\ i \neq 2m}}^{L-1} \left( \ket{i} \bra{i+1} + \text{h.c.} \right)  - \gamma_2 \left( \ket{2m} \bra{2m+1} + \text{h.c.} \right) \\
        & - i\gamma_2 \left( \ket{2m} \bra{2m} - \ket{2m+1} \bra{2m+1} \right) \\
        & - i\gamma_1 \ket{m} \bra{m} - i\gamma_3 \ket{3m+1} \bra{3m+1}
\end{aligned}
\end{equation}
This Hamiltonian has PC at $\gamma_1=\gamma_3=2$, $\gamma_2=1$, as demonstrated in Figure \ref{fig_higher_order}.   The number of visible eigenvalues shows a dramatic four-fold reduction at this point.

The analytic proof of pairwise coalescence in Section \ref{sec:mainproof} can be extended to the case of this phenomenon of full-spectrum higher-order coalescences, by showing that the characteristic polynomial $P_L(\lambda)$ of the Hamiltonian can be expressed as $[F(\lambda)]^4$, where $F(\lambda)$ is some polynomial.   Also, the scheme can be repeated iteratively to produce 8-fold, 16-fold,.... etc full-spectrum coalescences.  These developments are beyond the scope of this article and will be the subject of future work.


\section{Proof of Pairwise Coalescence under off-Center reflection symmetry \label{sec:mainproof}}

In this section, we will prove our main result announced in Section  \ref{sec:mainresult}, namely, that a tight-binding chain with off-center reflection symmetry has PC when the central $2\times2$ block has an exceptional point.  

After setting up the problem in \S\ref{subsec:mainproof:setup}, we show in \S\ref{subsec:mainproof_squareform} that the eigenvalues of our Hamiltonian, Eq.\ \eqref{eq:Ham_general} with condition \eqref{eq:central_block_condition}, are all paired.  In \S\ref{subsec:mainproof_eigstates} we explain why this means that the eigenstates also coalesce in pairs, i.e., that the phenomenon is not just degeneracies but $L/2$ EPs.  In \S\ref{subsec:generalizing_central_block} we point out a generalization that is permissible, but which we avoid using in the rest of the article for notational convenience.

\subsection{Setup \label{subsec:mainproof:setup}}

We will use the following recursion relation known for the determinant of tridiagonal matrices.   If the diagonals are $\mu _n$, the sub-diagonals and $\nu_n$ and the super-diagonals are $\sigma _n$, then 
\begin{equation}
\label{Trid}
    P_n=\mu_n P_{n-1}-\sigma_{n-1}\nu_{n-1}P_{n-2}
\end{equation}
where $P_n$ is the determinant of the $n\times n$ matrix formed by taking the first $n$ rows and first $n$ columns of the tridiagonal matrix.  In other words, $P_n$ is the determinant of the $n\times{n}$ matrix with elements
\begin{equation}
 \mu_i\delta_{i,j+1}+\nu_{j}\delta_{i+1,j}+ \sigma_i\delta_{ij}. 
\end{equation}
In addition, we define $P_0=1$ and $P_{-1}=0$. 
The recurrence relation \eqref{Trid} can be verified by Laplace expansion of the determinant along the last row or last column.  It appears in numerous sources, e.g., it is mentioned in Section 8.4 of Ref.~\cite{Golub_VanLoan_MatrixComputations_book1996}.  It will also be useful to write the recurrence relation as a matrix equation 
\begin{equation}  \label{eq:recurrence_matrixform}
\begin{pmatrix}P_n\\ P_{n-1}\end{pmatrix} = 
\begin{pmatrix}\mu_n & -\sigma_{n-1}\nu_{n-1}\\ 1 & 0\end{pmatrix} 
\begin{pmatrix}P_{n-1}\\ P_{n-2}\end{pmatrix} .   
\end{equation}
We wish to analyze the tight-binding Hamiltonian of Eq.\ \eqref{eq:Ham_general} with $L=2k$ sites, with the condition $\delta=(\gamma-\alpha)/2$.  To infer  the structure of the eigenvalues, we examine the characteristic polynomial, which (up to a minus sign) is given by the determinant  
\begin{widetext}
\begin{equation} \label{eq:charpol_matrix}
\det(\lambda I - H) = \det 
\begin{pmatrix}
\lambda+a_1 & b_1 & 0 & \cdots & 0 & 0 & 0 & 0 & 0 & 0 \\
c_1 & \lambda+a_2 & b_2 & \ddots & \vdots & \vdots & \vdots & \vdots & \vdots & \vdots \\
0 & c_2 & \lambda+a_3 & \ddots & 0 & 0 & 0 & 0 & \vdots & \vdots \\
\vdots & \ddots & \ddots & \ddots & b_{k-1} & 0 & 0 & 0 & \vdots & \vdots \\
0 & \cdots & 0 & c_{k-1} & \lambda+i\alpha & \frac{\gamma-\alpha}{2} & 0 & 0 & \vdots & \vdots \\
0 & \cdots & 0 & 0 & \frac{\gamma-\alpha}{2} & \lambda+i\gamma & b_{k-1} & 0 & \vdots & \vdots \\
0 & \cdots & 0 & 0 & 0 & c_{k-1} & \lambda+a_{k-1} & b_{k-2} & 0 & \vdots \\
0 & \cdots & 0 & 0 & 0 & 0 & c_{k-2} & \ddots & \ddots & 0 \\
\vdots & \vdots & \vdots & \vdots & \vdots & \vdots & 0 & \ddots & \lambda+a_2 & b_1 \\
0 & 0 & 0 & 0 & 0 & 0 & 0 & \cdots & c_1 & \lambda+a_1
\end{pmatrix} . 
\end{equation}
We have used $\lambda{I}-H$ instead of the conventional $H-\lambda{I}$, but since $L$ is even, this has no effect on the determinant.  

We have used the `generalized' off-diagonal reflection chosen also in Eq.\ \eqref{eq:Ham_general}: the $b_i$'s appear on the upper subdiagonal on both the first half and last half of the matrix.  If we wanted exact off-diagonal reflection symmetry, we would need to swap $b_n \leftrightarrow c_n$ for either the left or the right half.  In our proof of PC below, the $b$'s and $c$'s do not appear separately but only as products $b_nc_n$, which is why they can be swapped while sustaining PC.  Also, we have used $\delta=\frac{\gamma-\alpha}{2}$ here: the proof remains exactly the same if we use $\delta=-\frac{\gamma-\alpha}{2}$. 

We denote as $P_n$ the determinant of the leading $n\times{n}$ block of the matrix appearing in Eq.\ \eqref{eq:charpol_matrix}.  Thus $P_L=P_{2k}$ is the characteristic polynomial.  We will prove that it has the form $P_L(\lambda) = \left[F(\lambda)\right]^2$, implying that each root is a double root, i.e., that the eigenvalues all pair up.  

\subsection{Square form of characteristic polynomial \label{subsec:mainproof_squareform}}

We first use the recurrence relations in the form Eq.\ \eqref{eq:recurrence_matrixform} repeatedly, to express $P_L=P_{2k}$ in terms of  $P_{k+1}$ and $P_{k}$.  For this part of the matrix, $\mu_n=a_{L-n+1}$, $\nu_n=c_{L-n}$ and $\sigma_n= b_{L-n}$.  We define $b_nc_n=\eta_n$ for conciseness.   We thus obtain 
\begin{equation} \label{eq:Tmatrix_1}
\begin{pmatrix}
P_{2k} \\
P_{2k-1}
\end{pmatrix}
= \begin{pmatrix}
\lambda + a_1 & -\eta_1 \\
1 & 0
\end{pmatrix}
\begin{pmatrix}
P_{2k-1} \\
P_{2k-2}
\end{pmatrix}   
\\
= \begin{pmatrix}
\lambda + a_1 & -\eta_1 \\
1 & 0
\end{pmatrix}
\cdots
\begin{pmatrix}
\lambda + a_{k-1} & -\eta_{k-1} \\
1 & 0
\end{pmatrix}
\begin{pmatrix}
P_{k+1} \\
P_k
\end{pmatrix} = 
\begin{pmatrix}
T_{11} & T_{12} \\
T_{21} & T_{22}
\end{pmatrix}
\begin{pmatrix}
P_{k+1} \\
P_k
\end{pmatrix}. 
\end{equation}
We have defined the product of $k-1$ matrices as the matrix $T$.  

To make use the off-center reflection symmetry, we now write down the analogous relationship for the first $k-1$ rows and columns, using Eq.\ \eqref{eq:recurrence_matrixform} repeatedly to express $\begin{pmatrix}P_{k-1}\\ P_{k-2}\end{pmatrix}$ in terms of $\begin{pmatrix}P_{1}\\ P_{-1}\end{pmatrix}=\begin{pmatrix}1\\ 0\end{pmatrix}$.  It is convenient to write this in transposed form:
\begin{multline}  \label{eq:Amatrix_1}
\begin{pmatrix}
P_{k-1} &  
P_{k-2}
\end{pmatrix}
= \begin{pmatrix}
P_{k-2} & 
P_{k-3}
\end{pmatrix}    \begin{pmatrix}
\lambda + a_{k-1} & 1  \\
-\eta_{k-2} & 0
\end{pmatrix}
 =  \begin{pmatrix}
P_{0} & 
P_{-1}
\end{pmatrix}    
\begin{pmatrix} \lambda + a_{1} & 1  \\ -\eta_0 & 0 \end{pmatrix}
\begin{pmatrix} \lambda + a_{2} & 1  \\ -\eta_1 & 0 \end{pmatrix}
\cdots \begin{pmatrix} \lambda + a_{k-1} & 1  \\ -\eta_{k-2} & 0 \end{pmatrix}
\\ =  \begin{pmatrix} 1&  0 \end{pmatrix}    
\begin{pmatrix} \lambda + a_{1} & 1  \\ 0 & 0 \end{pmatrix}
\begin{pmatrix} \lambda + a_{2} & 1  \\ -\eta_1 & 0 \end{pmatrix}
\cdots \begin{pmatrix} \lambda + a_{k-1} & 1  \\ -\eta_{k-2} & 0 \end{pmatrix}
 =  \begin{pmatrix} 1&  0 \end{pmatrix}    
\begin{pmatrix} A_{11} & A_{12}  \\ A_{21} & A_{22} \end{pmatrix}
. 
\end{multline}

\end{widetext}
The $A$ matrix defined here is, like the $T$ matrix, also a product of $k-1$ matrices.  The component matrices of $A$ are composed of almost the same elements as the matrices making up the $T$ matrix in Eq.\ \eqref{eq:Tmatrix_1}; this is a manifestation of the off-diagonal reflection symmetry.  We therefore expect the $T$ and $A$ matrices to be closely related.  Indeed, examining the matrices for $k=2$ and $k=3$ reveals 
\begin{equation}  \label{eq:A_and_T}
A_{11} = T_{11} , \quad  A_{12} = -\frac{T_{12}}{\eta_{k-1}}     
\end{equation}
for these cases.  By examining the definitions of $A$ and $T$, one can show   that these relations hold for any integer $k\geq2$.   The derivation of Eqs.\ \eqref{eq:A_and_T} is delegated to Appendix \ref{appsec:derivation_ATrelation}.  From Eqs.\ \eqref{eq:Amatrix_1} and \eqref{eq:A_and_T}, we obtain 
\begin{gather}
P_{k-1} = A_{11} = T_{11} , \\
P_{k-2} = A_{12} = - T_{12}/ \eta_{k-1} .  
\end{gather}
Combining these results with Eq.\ \eqref{eq:Tmatrix_1} yields
\begin{multline}   \label{eq:PL_in_4_Ps}
P_{2k} = T_{11} P_{k+1} + T_{12} P_k 
\\ =  P_{k+1} P_{k-1} - \eta_{k-1} P_k P_{k-2} .   
\end{multline}
We have obtained Eq.\ \eqref{eq:PL_in_4_Ps} by using the structure of the matrix outside the central block, in the form of Eqs.\ \eqref{eq:Tmatrix_1} and \eqref{eq:Amatrix_1}.  We will now use the structure of the central block.  The recursion relations applied to this part of the matrix yield 
\begin{equation}
\label{eq:recursion_central_block}
    \begin{aligned}
       P_{k+1} &= (\lambda +i\gamma)P_k-\left(\frac{\gamma-\alpha}{2}\right)^2P_{k-1} , \\
       P_k &= (\lambda+i\alpha)P_{k-1}-\eta_{k-1}P_{k-2} . 
    \end{aligned}
\end{equation}
Using these, we eliminate $P_{k+1}$ and $P_{k-2}$ from Eq.\ \eqref{eq:PL_in_4_Ps}, obtaining $P_L=P_{2k}$ solely in terms of $P_{k}$ and $P_{k-1}$.  The result is 
\begin{equation}
\label{eq:PL_as_square}
    P_L=P_{2k}=\left(P_{k}+i\left(\frac{\gamma-\alpha}{2}\right)P_{k-1}\right)^2.
\end{equation}
We have thus proved that the characteristic polynomial has a squared form, which implies that all eigenvalues of the Hamiltonian are paired.

\subsection{Pairing of eigenvalues implies eigenstate coalescence \label{subsec:mainproof_eigstates}}

We have shown that the eigenvalues of our Hamiltonian are paired.  We will now argue that, for a Hamiltonian described by a tridiagonal matrix (such as ours), coalescence of eigenvalues implies a
coalescence of eigenstates, i.e., that the eigenstates corresponding to the equal eigenvalues are
always linearly dependent.   This will demonstrate that the pairs with identical eigenvalues are in fact exceptional points and not just degeneracies.   

We expand on a version of the argument that has appeared in the Appendix of Ref.\ \cite{Burke2020} in a similar context.  
Consider an eigenvalue $\lambda$ of our tridiagonal matrix $H$, and corresponding eigenvector $\ket{Z}=(z_1,z_1,\ldots,z_L)^T$.  The eigenvalue equation $H\ket{Z}=\lambda\ket{Z}$ is a system of $L$ linear equations.  Due to the tridiagonal form of the matrix $H$, we can `solve' for $z_2$, $z_3$, $\ldots$ in terms of $z_1$: 
\begin{equation}
z_2 = \xi_2 z_1, \; z_3 = \xi_3 z_1, \; 
z_4 = \xi_4 z_1, \; \ldots     
\end{equation}
with the constants $\xi_i$ depending only on $\lambda$ and the elements of $H$.  This works as long as the tridiagonal Hamiltonian is not diagonal, i.e., at least one of the sub- or super-diagonals is nonzero.  

If a second eigenvector corresponds to the same eigenvalue  $\lambda$ (coalescing eigenvalues), then the $\xi_i$ are the same for the two eigenvectors, i.e., ratios $z_i/z_1$ are identical for the two eigenvectors.  Thus, the two eigenvectors can only deviate by a constant factor, i.e., they are linearly independent.

\subsection{Generalizing the central block \label{subsec:generalizing_central_block}}

We have sacrificed some generality by restricting the central block to have the form of Eq.\  \eqref{eq_centralblock_Ham}, i.e., real and symmetric hopping on central bond, purely imaginary potentials on central sites.  Then, we have $3$ real parameters to describe the central block, with one PC condition \eqref{eq:central_block_condition} connecting them, thus leaving us with a two-parameter family after the rest of the chain has been specified.  

This restriction simplified our description, but is not necessary for PC.  More generally, we could have used a central block of the form 
\begin{equation}
M = \begin{bmatrix}
    H_{k,k} & H_{k,k+1} \\
    H_{k+1,k} & H_{k+1,k+1}
\end{bmatrix}
= - \begin{bmatrix}
    i\tilde{\alpha} & \delta_1 \\
    \delta_2 & i\tilde{\gamma}
\end{bmatrix} ,   
\end{equation}
where $\tilde{\alpha}$, $\tilde{\gamma}$, $\delta_1$ and $\delta_2$ are now complex.  The requirement of an EP (requiring that the characteristic polynomial is a square) imposes the condition 
\begin{equation}  \label{eq:central_block_condition_generalized}
    \delta_1\delta_2 = \frac{1}{4} \left(\tilde{\alpha}-\tilde{\gamma}\right)^2 
\end{equation}
which is the generalization of Eq.\ \eqref{eq:central_block_condition}.  

The proof of PC provided in this section can be readily extended to this more general family of central blocks.  
With 4 complex parameters and one constraint, the central block actually provides a $7$-paramater family of Hamiltonians with PC, after the rest of the chain is fixed.


\section{Non-Orthogonality of Eigenstates\label{sec:nonorthogonal}}

The eigenvectors of Hermitian matrices are orthogonal and there is no distinction between left and right eigenvectors. In contrast, in non-Hermitian quantum mechanics, left and right eigenvectors are generically different.  It is conventional to focus on the right eigenvectors; this is what we will do in this work.   

Unlike the Hermitian case, the eigenvectors of non-Hermitian matrices are not orthogonal.  This non-orthogonality is acute at exceptional points because eigenvectors become linearly dependent.  In this paper we constructed Hamiltonians where the full spectrum breaks up into pairs of coalescing pairs; one could expect this to have a dramatic effect on non-orthogonality.  It is therefore interesting to examine the systems we have proposed through the lens of non-orthogonality.  

To have a measure of the non-orthogonality of right eigenstates, we form the matrix with elements $U_{\mu,\nu}=\langle \mu | \nu \rangle$ where $\mu, \nu$ are eigenstate indices and $\ket{\mu}$, $\ket{\nu}$ are normalized right eigenstates.   Here $\bra{\mu}$ represents the hermitian conjugate of the right eigenvector $\ket{\mu}$  (and does not represent any left eigenvector).  

\begin{figure}[tbp]
    \centering
    \includegraphics[width=8.6cm]{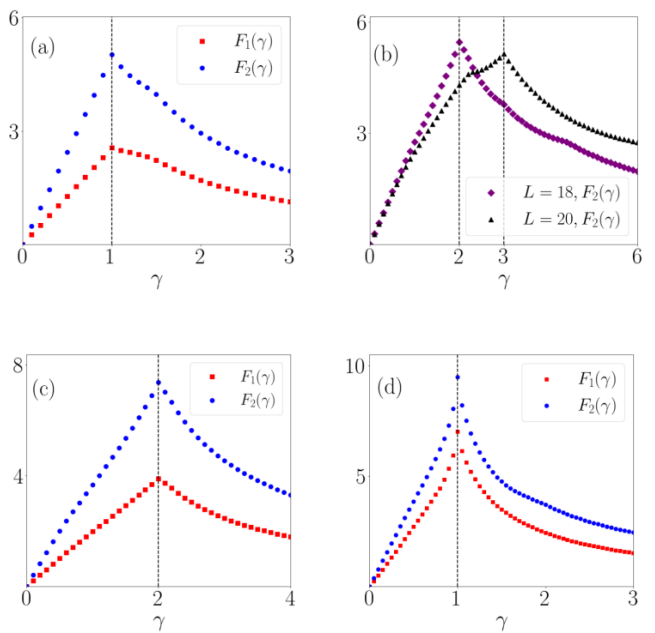}
    \caption{PC maximizes the non-orthogonality of eigenstates in all families. The quantities defined in \eqref{F_1} and \eqref{F_2} are maximized at the $\gamma$ value that leads to PC. (a) In Family A, we choose $\alpha=0$ and $\delta=0.5$, so that PC is at $\gamma=1$.  (b) For Family B we plot only $F_2(\gamma)$ for two different system sizes, with $\alpha=0$.  We choose $J_1=1, J_2=1.5$, for both $L$.  $F_2$ is maximum when there is PC. (c) In Family C, we choose $J_1=J_2=1.5$ and $J_3=1$ with $L=30$. (d) For Family D, we choose $\gamma_1/2=\gamma_3/2=\gamma_2=\gamma$ and all the hoppings are 1 with $L=20$. 4-fold coalescence happens at $\gamma=1.$ }
    \label{fig:sumsquares}
\end{figure}

Since we use normalized right eigenvectors, the diagonal terms of $U$ are trivially equal to one ($\langle\mu|\mu\rangle =1$), so the interesting terms are the off-diagonal ones.  We therefore subtract the $L\times L$ identity matrix from $U$ and consider the matrix $U-I$.  The norm of this matrix quantifies the non-orthogonality of the system.   In Figure \ref{fig:sumsquares}, such norms are plotted as a function of $\gamma$, with an example each from Family A, B, C and D.  In each case, a sharp peak is seen at the value of $\gamma$ where PC occurs.  Clearly, the PC phenomenon very substantially enhances non-orthogonality.

The quantities plotted in Figure \ref{fig:sumsquares} are 
\begin{align}
    F_1(\gamma) &=\frac{1}{L} \sum_{\mu=1}^L \sum_{\nu=1}^L |(U-I)_{\mu,\nu}|
    \label{F_1}
    \\
    F_2(\gamma) &= \sqrt{\sum_{\mu=1}^L \sum_{\nu=1}^L \Big[(U-I)_{\mu,\nu}\Big]^2}
    \label{F_2}
\end{align}
where $(U-I)_{\mu,\nu}$ is the $(\mu,\nu)$ entry of the matrix $U-I$.  The second quantity $F_2$ is the Hilbert-Schmidt norm (also known as the Frobenius norm or Schatten 2-norm or Euclidean norm).  Another common matrix norm, the trace norm or Schatten 1-norm, is not useful in this case as $U-I$ is always traceless by construction.  We have added $F_1$, another measure of the largeness of the elements of $U-I$, in order to make sure that the dramatic enhancement of non-orthogonality at the PC point is not an artifact of how we characterize the matrix.

\begin{figure}[tbp]
    \centering
    \includegraphics[width=\linewidth]{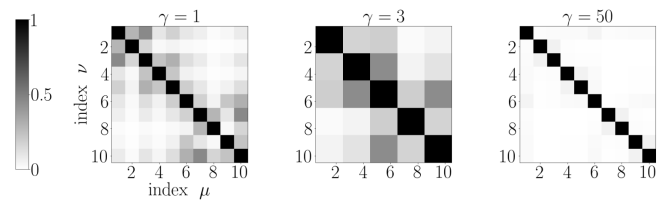}
    \caption{Complex norm of the matrix elements of $U_{\mu,\nu}=\langle \mu | \nu \rangle$ for Family B. We see that when PC occurs (middle panel), successive eigenvectors are linearly dependent. At large $\gamma$ (right panel), the non-orthogonality matrix becomes almost diagonal, so that eigenstates become nearly orthogonal. Here $L=10, J_1=1.5, J_2=1$ and $\alpha=0$, so that PC occurs at $\gamma=2J_1=3$.}
    \label{Non_Ortho_Color_Plot}
\end{figure}

In Figure \ref{Non_Ortho_Color_Plot}, we show the magnitudes of all the matrix elements of the non-orthogonality matrix $U$, for an example from Family B.   Since the eigenvalues lie on the complex plane, there is no natural ordering of eigenvalue indices $\mu$ and $\nu$.  One possibility is to first order the eigenvalues by real parts and then by imaginary parts and we use this method here.  With this ordering, the coalescing eigenstates have consecutive indices. 

The middle panel ($\gamma=3$) corresponds to PC.  The coalescence of eigenstates is clearly visible in the non-orthogonality matrix: coalescing pairs of eigenstates have unit overlap, so there are $2\times2$ blocks of $1$'s.  In the other cases, only the diagonal elements are equal to $1$.

At large  $\gamma$ values, for all families,  the off-diagonal entries of $U$ are vanishingly small.  We could interpret this as a quantum Zeno effect: the Hamiltonians behave effectively  `Hermitian' when the dissipative impurity is too large.  In Figure \ref{fig:sumsquares}, this is reflected in $F_1$ and $F_2$ going to zero for $\gamma\to\infty$.  In Figure \ref{Non_Ortho_Color_Plot},  the smallness of off-diagonal elements can be seen visually in the right panel ($\gamma=50$).

\section{Enhanced loss of norm\label{sec:normloss}}

Notable features in the spectrum can be expected to have dynamical consequences.  We speculate that PC --- arrangement of the full spectrum into coalescent pairs --- will have several types of dynamical signatures.  A detailed study of various types of dynamics, for the diverse class of Hamiltonians showing PC that we have presented, is beyond the scope of this paper.  In this section we present one dynamical effect, related to the dynamics of the norm of the wavefunction.

A fundamental feature of non-Hermitian Hamiltonians is that the wavefunction norm is not preserved under time evolution.  For an absorbing imaginary potential (such as $-i\gamma$ with positive $\gamma$), the norm generally decreases with time.  We have found that the loss of norm is enhanced by the PC phenomenon.  

\begin{figure}[tbp]
\centering
\includegraphics[width=\columnwidth]{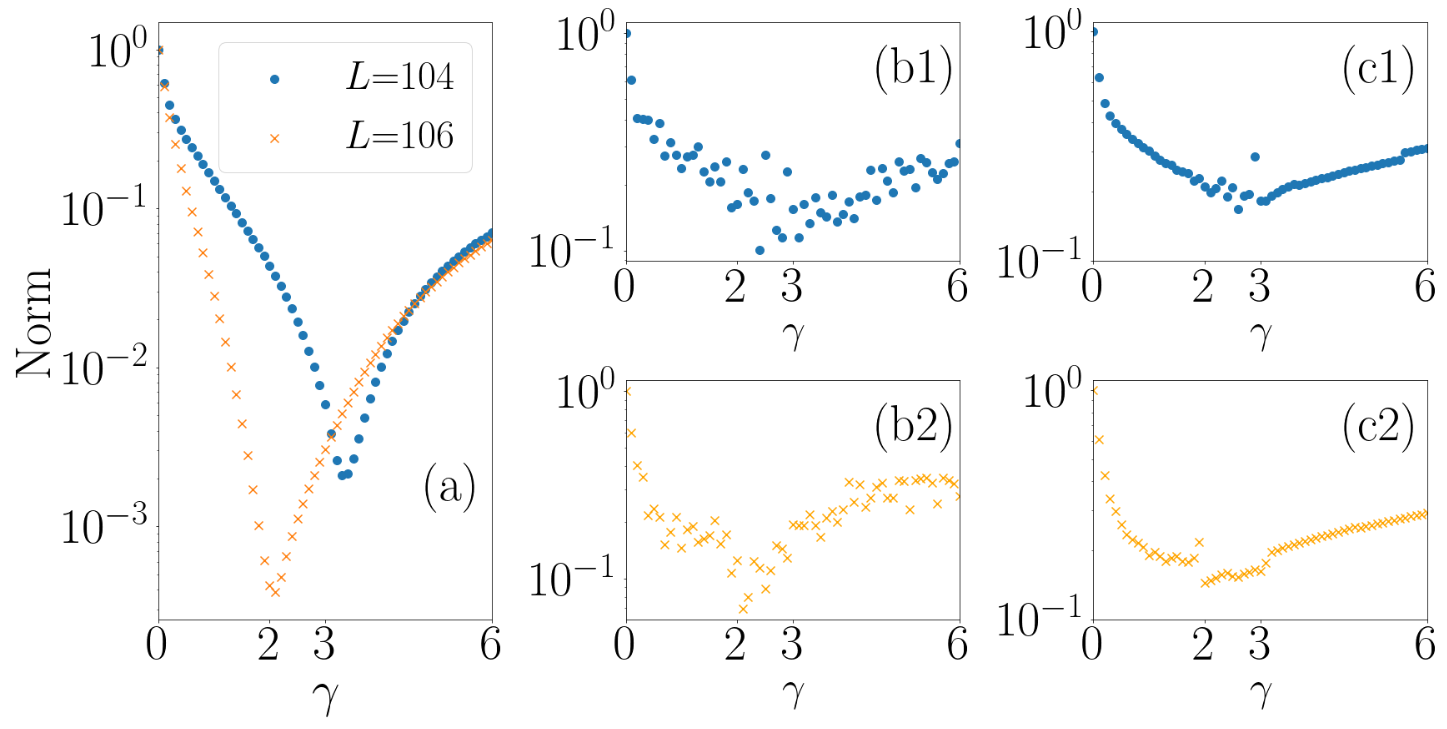} 
\caption{\label{fig:norm_after_sometime_vs_gamma} The value of $\gamma$  for minimum norm (after evolution for time $t=3L$ in all cases) and for PC are close. (a) Norm of a wavepacket with  $j_0=L/4, \sigma=L/8, J_1=1, J_2=1.5$ at $t=3L$ with $p=\pi/4$, after evolution for time $t=3L$.  The minimum of the norm is close to the value where PC occurs (at $\gamma=2J_1=2$ for $L=4m+2$ and at $\gamma=2J_2=3$ for $L=4m$).  
(b) Initial wave function is uniformly distributed in the lattice: $d_j=1/\sqrt{L}$ for all $j$.
(c) Similar phenomenon is seen when initial wave function is uniformly distributed among eigenstates: $c_\mu=1/\sqrt{L}$ for all $\mu$. Panels (b1), (c1) are for $L=104$ and (b2), (c2) are for $L=106$.}
\end{figure}

We consider the SSH model with a single central onsite potential ($\alpha=0$).  The norm after a relatively long time, $t=3L$, is shown in Figure \ref{fig:norm_after_sometime_vs_gamma} as a function of $\gamma$, for several different initial states.  In each case, the decrease of the norm is seen to be more drastic when $\gamma$ is near the PC point, i.e., the norm vs $\gamma$ curves have minima near $\gamma=2J_1$ or $\gamma=2J_2$ depending on whether we have $L=4m+2$ or $L=4m$.   The correspondence is close but not exact. 

In panel (a), we show the norm in wave packet dynamics.  We  start with the initial wavefunction 
\begin{equation} \label{eq:expand_initial_state_in_site_basis}
\ket{\psi(0)} = \sum _{j=1}^L d_j \ket{j}
\end{equation}
where $j$ is the site index, with coefficients  
\begin{equation}   \label{eq:gaussian_wavepacket_coeffs}
d_j = \exp\left[-\frac{(j-j_0)^2}{2\sigma^2}\right]e^{ipj} . 
\end{equation}
This is a Gaussian-shaped wave packet centered at site $j_0$ and having width $\sigma$ and momentum $p$.  The momentum determines the speed of the wavepacket.  The wavepacket is sent toward the imaginary potential from $j_0=L/4$, and multiple scattering events are allowed to happen before measuring the norm at $t=3L$.  During this time, the wavepacket splits into multiple pieces and disperses, and also has other complicated dynamical features due to the two-band nature of the underlying lattice.   However, the behavior of the norm after time $t=3L$ as a function of $\gamma$ is relatively simple and shows clear minima quite close to the PC point in each case.  

In panel (b), we use an initial state uniformly spread in space, i.e., coefficients $d_j=1/\sqrt{L}$ in Eq.\ \eqref{eq:expand_initial_state_in_site_basis}.  In panel (c) we use an initial state uniformly distributed in the eigenstate basis, i.e., expressing the initial state as 
\begin{equation} \label{eq:expand_initial_state_in_eigenstate_basis}
\ket{\psi(0)} = \sum _{\mu=1}^L c_\mu \ket{\mu}
\end{equation}
where $\mu$ is the eigenstate index, we choose for panel (c) the coefficients $c_\mu=1/\sqrt{L}$.    In these cases, the norm at a fixed time is a noisier function of $\gamma$ compared to the wavepacket initial state in panel (a), however, a minimum near the PC point is clear in each case.   

To explain this phenomenon, we use the expansion of Eq.\ \eqref{eq:expand_initial_state_in_eigenstate_basis} to find the state vector at a later time: 
\begin{equation}
\ket{\psi(t)} = e^{-iHt}\ket{\psi (0)} =  \sum _{\mu=1}^{L}c_\mu e^{-iE_\mu t}\ket{\mu}.    
\end{equation}
where $E_\mu$ is the corresponding eigenvalue.  From this, we can calculate the norm at an arbitrary time $t$:
\begin{equation}
N(t) = \langle \psi(t)|\psi(t)\rangle=\sum_{\mu,\nu}c_\mu^{*}c_\nu e^{it(E_\mu^{*}-E_\nu)}\langle \mu|\nu\rangle.
\label{eq:generalnormeq}
\end{equation}
Since the eigenbasis is not complete when the system has an exceptional point or PC in the system, there are situations for which the above expansion cannot be used.  However, this restriction will not play a role in the argument below.  Eq.\ \eqref{eq:generalnormeq} is not easy to analyze, so we consider the approximation $\langle \mu|\nu\rangle\approx \delta_{\mu,\nu}$, i.e., ignore the non-orthogonality:
\begin{equation}
N(t) \approx \sum_{\mu=1}^L e^{2tI_\mu}|c_\mu|^2 
= \sum_{\mu=1}^L e^{-2t\left|I_\mu\right|} |c_\mu|^2  ,  
\label{eq:approx_eq_for_norm}
\end{equation}
where $I_\mu$ is the imaginary part of the $\mu$-th eigenvalue, which is non-positive for a non-Hermiticity generated by an absorbing potential $-i\gamma$ with $\gamma>0$.   
If we consider any pair of eigenvalues, noting that  $2e^{-x} \leq e^{-x-\delta} + e^{-x+\delta}$ for positive $x$ and $\delta$, we see that the approximate expression is minimized when the eigenvalues coincide.  With PC, the complete spectrum is paired up, so that this effect is enhanced.  Because Eq.\ \eqref{eq:approx_eq_for_norm} is approximate, this argument is only approximately valid: this explains why $N(t)$ generally has a minimum near but not exactly at PC.

\section{Discussion and Context\label{sec:discussion}}

We have presented a scheme for generating non-Hermitian tight-binding Hamiltonians with the remarkable property of pairwise coalescence, such that the \emph{complete} spectrum pairs up.  This is a significantly enhanced version of a regular exceptional point where a single pair of eigenstates coalesces.  We have studied a few families of models designed in this manner.  By recursively applying the idea, one can also devise Hamiltonians with enhanced versions of 4th-order or even higher-order EPs.  

The symmetries that generate PC involve reflection omitting the central two sites, plus exchanges of the rightward and leftward hopping amplitudes on any number of the bonds.  In Eqs.\ \eqref{eq:Ham_general} and \eqref{eq:charpol_matrix} we have chosen one particular symmetry --- rightward and leftward amplitudes exchanged on every bond of the right half of the chain.  But this was only for notational definiteness; the scheme allows one to choose from a whole class of symmetry operations.  Technically, this is because the coefficients $b_n$ and $c_n$ only appear in the combination $b_nc_n$ in the proof of Section \ref{sec:mainproof}.  Similarly, although we have focused on symmetric real hopping in the central bond and imaginary potentials on the central sites, the actual choice is much larger, as explained in \S\ref{subsec:generalizing_central_block}.

Partly because of this generality, a large number of different types of Hamiltonians are encompassed by our setup.  We have explored only a few, mainly for demonstration purposes, e.g., a highly random case (Figure \ref{fig_random_matrixelements}), a very regular case containing a uniform lattice everywhere except the central two sites and central bond (Family A), SSH chains with imaginary potentials at one or both of the central sites (Family B), and the Hamiltonians with 4-wise coalescence of the complete spectrum (Family D).  Clearly, many other types of physics can be incorporated.  For example, with non-reciprocal hoppings, one can still preserve our generalized off-center reflection symmetry.  This opens the intriguing possibility of the interplay between PC and the non-Hermitian skin effects \cite{Lee_PRL2016_AnomalousEdgeState, Yao_Wang_PRL2018_EdgeStates_TopologicalInvariants, MartinezAlvarez_BarriosVarga_FoaTorres_PRB2018_robustedgestates}.  

Exceptional points have generated interest partly because of the possibilities they might offer for control of optical and other dissipative or non-reciprocal systems.  Since PC extends the coalescence phenomenon to the full spectrum, this enhanced version should also allow amplification of various consequences of EPs.  As an example, there is the idea of using EPs for sensing: the sensitivity might be multiplied manyfold if we have many pairs rather than a single pair of coalescing states.  This possibility becomes even more promising when we make the spectrum coalesce in groups of 4 or 8 or $2^n$, as can be done by repeatedly applying our scheme to a large tight-binding chain.   

Our work opens up various research questions and directions, in addition to the possible extensions and applications pointed out above.  
We have made some surface explorations of the effects of PC in this paper, but it is likely that these studies are only scratching the surface.   For example, exploring the effect of PC on time evolution in detail remains an open task.   It would also be interesting to look for modifications and generalizations of our off-diagonal reflection symmetry for other lattice geometries, for example, ladders and two-dimensional lattices.  This might lead to the interplay between non-Hermitian flat-band physics \cite{Leykam_Flach_Chong_PRB2017_flatbands, Maimaiti_Andreanov_PRB2021_flatband} and PC, or between two-dimensional skin effects and PC.


\section*{Acknowledgments}
MH acknowledges financial support from the Deutsche Forschungsgemeinschaft under grant SFB 1143 (project-id 247310070).


\appendix


\section{Derivation of Eq.\ \eqref{eq:A_and_T} \label{appsec:derivation_ATrelation}}

In this Appendix we outline the derivation of Eq.\ \eqref{eq:A_and_T} for arbitrary $k\geq2$.   

The $A$ and $T$ matrices were defined as products of $2\times2$ matrices,  $A=A^{(1)}A^{(2)}...A^{(k-1)}$ and $T=T^{(1)}T^{(2)}...T^{(k-1)}$, where 
\begin{equation}
A^{(i)}=
\begin{pmatrix}
\lambda + a_{i} & 1 \\
-\eta _{i-1} & 0
\end{pmatrix} ,  \quad 
T^{(i)}=
\begin{pmatrix}
\lambda + a_{i} & -\eta_i \\
1 & 0
\end{pmatrix},     
\end{equation}
with $\eta_0 := 0$.  We will use mathematical induction to show that Eq.\ \eqref{eq:A_and_T}, i.e.  the relations $A_{11}=T_{11}$ and $A_{12}=-T_{12}/\eta_{k-1}$,  hold for arbitrary integer $k\geq2$.  To do this, we show [A] that  Eq.\ \eqref{eq:A_and_T} holds for $k=2$, and [B] if it holds for $k=i$, then it also holds for $k=i+1$.  

[A] For $k=2$, 
\[
A = A^{(1)}=
\begin{pmatrix}
\lambda + a_{1} & 1 \\
-\eta _{0} & 0
\end{pmatrix} ,  \quad 
T=T^{(1)}=
\begin{pmatrix}
\lambda + a_{1} & -\eta_1 \\
1 & 0
\end{pmatrix}, 
\]
so that $A_{11} =\lambda+a_{1} =T_{11}$ and $A_{12}=1 = -T_{12}/\eta_{1}= -T_{12}/\eta_{k-1}$, i.e., Eq.\ \eqref{eq:A_and_T} holds.  This completes the first step of mathematical induction. 

[B] If Eq.\ \eqref{eq:A_and_T} holds for $k=i$, then the matrices $A$ and $T$ can be parametrized as 
\begin{equation}
A=A^{(1)}A^{(2)}...A^{(i-1)} =
\begin{pmatrix}
x_1 & x_2 \\
x_3 & x_4
\end{pmatrix}    
\end{equation}
and 
\begin{equation}
T=T^{(1)}T^{(2)}...T^{(i-1)} =
\begin{pmatrix}
x_1 & - x_2 \eta_{i-1} \\
x_5 & x_6
\end{pmatrix} .     
\end{equation}
To obtain the $A$ and $T$ matrices for $k=i+1$, we should multiply the first matrix by $A^{(i)}$ and the second matrix by $T^{(i)}$.   The results are 
\begin{multline}
A=A^{(1)}A^{(2)}...A^{(i-1)}A^{(i)} 
 = \begin{pmatrix}
x_1 & x_2 \\
x_3 & x_4
\end{pmatrix}  
\begin{pmatrix}
\lambda + a_{i} & 1 \\
-\eta _{i-1} & 0
\end{pmatrix}
\\ = \begin{pmatrix}
(\lambda+a_{i})x_1-x_2\eta_{i-1} & x_1 \\
(\lambda+a_{i})x_3-x_4\eta_{i-1} & x_3
\end{pmatrix}    
\end{multline}
and 
\begin{multline}
T=T^{(1)}T^{(2)}...T^{(i-1)}T^{(i)} 
\\ = \begin{pmatrix}
x_1 & - x_2 \eta_{i-1} \\
x_5 & x_6
\end{pmatrix} 
\begin{pmatrix}
\lambda + a_{i} & -\eta_i \\
1 & 0
\end{pmatrix}
\\ =
\begin{pmatrix}
(\lambda+a_{i})x_1-x_2\eta_{i-1} & -x_1\eta_{i} \\
(\lambda+a_{i})x_5-x_6\eta_{i-1} & -x_5\eta_{i}
\end{pmatrix} .     
\end{multline}
Comparing the two matrices, we see that $A_{11} =T_{11}$ and $A_{12}= -T_{12}/\eta_{i}$, i.e., Eq.\ \eqref{eq:A_and_T} holds for $k=i+1$ if it holds for $k=i$.  This completes the second step of mathematical induction.

\bibliography{bibbo}

\end{document}